\documentstyle[12pt,epsfig,aaspp4]{article}

\newcommand\Kms{$\rm km\,s^{-1}$}
\newcommand\nop{\noindent} 

\begin{document}

\title{Inverse Compton Scattering in Mildly Relativistic Plasma}

\author{S. M. Molnar\altaffilmark{1,2}}
\affil{Laboratory for High Energy Astrophysics, Code 662 \\
                   Goddard Space Flight Center \\
                      Greeenbelt, MD 20771}

\author{M. Birkinshaw\altaffilmark{3}}
\affil{Department of Physics, University of Bristol, \\
    Tyndall Avenue, Bristol, BS8 1TL, UK}

\altaffiltext{1}{NAS/NRC Research Associate}

\altaffiltext{2}{previous address: Department of Physics, University of Bristol,
                                   Tyndall Avenue, Bristol, BS8 1TL, UK}

\altaffiltext{3}{also: Center for Astrophysics,
    60 Garden Street, Cambridge, MA 02138, USA}

\begin{abstract}

We investigated the effect of inverse Compton scattering in mildly relativistic
static and moving plasmas with low optical depth using Monte Carlo simulations, 
and calculated the Sunyaev-Zel'dovich effect in the cosmic background radiation. 
Our semi-analytic method is based on a separation of photon diffusion in 
frequency and real space. We use Monte Carlo simulation to derive the intensity 
and frequency of the scattered photons for a monochromatic incoming radiation.
The outgoing spectrum is determined by integrating over the spectrum of the 
incoming radiation using the intensity to determine the correct weight. This 
method makes it possible to study the emerging radiation as a function of 
frequency and direction. As a first application we have studied the effects of 
finite optical depth and gas infall on the Sunyaev-Zel'dovich effect (not 
possible with the extended Kompaneets equation) and discuss the parameter range 
in which the Boltzmann equation and its expansions can be used. For high 
temperature clusters ($k_B T_e \gtrsim 15$ keV) relativistic corrections based 
on a fifth order expansion of the extended Kompaneets equation seriously 
underestimate the Sunyaev-Zel'dovich effect at high frequencies. The 
contribution from plasma infall is less important for reasonable velocities. We 
give a convenient analytical expression for the dependence of the cross-over 
frequency on temperature, optical depth, and gas infall speed. Optical depth 
effects are often more important than relativistic corrections, and should be 
taken into account for high-precision work, but are smaller than the typical 
kinematic effect from cluster radial velocities.

\end{abstract}

Subject headings: 
(cosmology:) cosmic microwave background --- 
galaxies: clusters: general --- 
methods: numerical --- 
plasmas --- 
scattering

% = = = = = = = = = = = = = = = = = = = = = = = = = = = = = = = = = = = = = = = = = = = = = = = = = =
\section{Introduction}
% = = = = = = = = = = = = = = = = = = = = = = = = = = = = = = = = = = = = = = = = = = = = = = = = = =

Inverse Compton scattering of the cosmic microwave background radiation (CMBR) by hot electrons in 
the atmospheres of clusters of galaxies, the Sunyaev-Zel'dovich (SZ) effect (Sunyaev and
Zel'dovich 1980), has become a powerful tool in astrophysics. It is one of the
most important secondary effects which cause fluctuations in the CMBR.
We will refer to the effect arising from static gas as the static 
SZ (SSZ) effect, and that arising from gas with bulk motion as the kinematic SZ (KSZ) effect. 
Fluctuations in the CMBR caused by the SZ effects in an ensemble of clusters of galaxies
should dominate on angular scales less than few arc minutes. The nature of these 
fluctuations depends on the evolution of clusters, and so it is a test of structure formation 
theories (eg. \cite{Aghanimet98};  \cite{MolnarBirkinshaw98}). 

Observations of the SSZ effect, begun in the 1970s, have now become routine 
in the 90s with dedicated instruments using the latest receiver technology 
(for reviews see \cite{Rephaeli95b}; \cite{Birkinshaw99}).
SZ effect and x-ray measurements probe the physical conditions in the intracluster 
gas in clusters of galaxies, and allow us to deduce the distance to the cluster without 
additional assumptions. This provides a useful independent method for determining the 
Hubble constant. 

Observations of the KSZ effect are much more difficult since it
is typically an order of magnitude smaller than the SSZ effect and has the same spectrum
as the primordial fluctuations in the CMBR. The KSZ effect provides a method of measuring radial 
peculiar velocities of clusters, and even without the tangential velocity component, 
which might be determined using the Rees-Sciama (RS) effect (\cite{ReesSciama68}; 
\cite{BirkinshawGull83}; \cite{GurvitsMitrof86}; \cite{Aghanimet98}; \cite{MolnarBirkinshaw98})
it should provide important information on large scale velocity fields, which are closely related 
to the large scale density distributions and thus to the average total mass density in the Universe.
Useful limits on the size of the KSZ effect for two clusters have recently been reported by
Holzapfel et al. (1997a).

Most discussions of the SSZ effect have been based on non-relativistic calculations of its 
amplitude, made via a Fokker-Planck type expansion of the Boltzmann equation 
(\cite{Kompaneets57}). 
The advantage of this approach is that in interesting cases
(electron temperature, $T_e$, much greater than the temperature of the incoming radiation)
it provides a convenient analytical solution for the spectrum of the emerging radiation. 
However, it has been recognized recently that relativistic effects become important for 
clusters with $k T_e \gtrsim 10$ keV.
Rephaeli (1995a) provided a relativistic solution for the SSZ effect as a series expansion 
in the optical depth ($\ll 1$ for clusters).
There is no exact analytical solution.
The numerical integrals involved are tractable in the single-scattering 
approximation, which is usually adequate in clusters. 
Fargion, Konoplich and Salis (1996) developed exact expressions for relativistic inverse 
Compton scattering of a laser beam with monochromatic isotropic radiation, finding good 
agreement with the approximations of Jones (1968). 
Their expression for the frequency redistribution function (FRDF) for scattering 
of mono-energetic electrons with monochromatic photons agrees with Rephaeli's result.
Rephaeli's method has been used in a series of papers to evaluate the SSZ effect for hot clusters. 
Relativistic corrections to the SSZ effect, and to the standard thermal bremsstrahlung formulae, 
were applied to determine the Hubble constant in hot clusters by 
Rephaeli and Yankovich (1997), however, their corrections of the thermal bremsstrahlung equation
were further corrected by Hughes and Birkinshaw (1998).
Holzapfel et al. (1997b) used the relativistic results in their determination of the Hubble constant 
from observations of cluster Abell 2163.

The most general treatment of Compton scattering in static and moving media has
been derived by Psaltis and Lamb (1997) as a series expansion.
As Challinor and Lasenby (1998b) noted, however, more terms in the expansion should be taken 
into account for accurate treatment of clusters of galaxies.
Recently the Kompaneets equation has been extended to contain relativistic corrections
to the SSZ and KSZ effects (\cite{Stebbins97}; \cite{Challinor98a}, b;
Itoh, Kohyama and Nozawa 1998; Nozawa, Itoh and Kohyama 1998; \cite{SazonovSun98b}). 
Starting from the Boltzmann equation, an expansion in the small parameters of the 
dimensionless temperature, $\Theta = k T_e /(m_e c^2)$, fractional energy change
in a scattering, $(h \nu^\prime - h \nu)/k_B T_e$,
and dimensionless radial velocity for the KSZ effect, $v_{rad}/c$, 
leads to a Fokker-Planck type equation (the extended Kompaneets equation).
Corrections up to the fifth order in $\Theta$ have been derived (\cite{Itohet98}).
These calculations demonstrate the importance of the relativistic effects (in accordance with
the results of Rephaeli 1995a). 
Note however, that, as Challinor and Lasenby (1998a) emphasized, 
the extended Kompaneets equation is a result of an asymptotic series expansion,
therefore it is important to estimate the validity of the expansion using other methods.
Nozawa et al. compared the convergence of their expansion to a direct numerical evaluation of
the Boltzmann collision integral, and concluded that in the Rayleigh-Jeans region the 
relativistic corrections give accurate results in the entire range of cluster temperatures. 
Significant deviations are found at higher frequencies for high temperature clusters.

The SSZ and KSZ effects must be separated in order to extract information on peculiar 
velocities. Fortunately the two effects have different frequency dependence.
The maximum of the KSZ effect (in thermodynamic temperature units) occurs at about the
``cross-over'' frequency where the SSZ effect changes sign from being a decrement to an increment.
In a non-relativistic treatment the cross-over frequency is a constant, 218 GHz, independent of
electron temperature, optical depth, and all other parameters.
Rephaeli (1995a) showed that in the relativistic case the cross-over frequency 
depends on the temperature, and his results were used
by Holzapfel et al. (1997a) in determining peculiar velocities of two clusters. 
Sazonov and Sunyaev (1998b) and Nozawa et al. (1998) give approximations for the 
cross-over frequency as a function of dimensionless temperature and radial peculiar velocity.
They also conclude that relativistic corrections to the cross-over frequency are important, 
and should be taken into account in future experiments.

Other methods have been used to investigate inverse Compton scattering, such as 
numerical integration of the collision integral (\cite{Corman70}), multiple scattering
methods (\cite{Wright79}), and Monte Carlo simulations.
Simulations of inverse Compton scattering in relativistic and non-relativistic plasma
have been carried out for embedded sources 
(Pozdnyakov, Sobol, and Sunyaev 1983; \cite{HaardtMaraschi93}; \cite{HuaTitarchuk95}).
Gull and Garret (1998) used Monte Carlo methods to evaluate the Boltzmann collisional
integral. Sazonov and Sunyaev (1998a) used Monte Carlo simulations to derive the SZ thermal and
kinematic effects.

% IN THIS PAPER

In this paper we study the effect of optical depth and non-uniform bulk motion on the 
SZ effect using a Monte Carlo method to calculate the frequency redistribution function.
The inverse Compton scattering of CMBR photons is treated in the Thomson limit for static 
and infalling plasmas (SSZ and KSZ effects) with spherical symmetry, uniform density distribution, 
and low optical depth over a wide range of gas temperatures and observed frequency. 
We apply our results to clusters of galaxies assuming a static and radially infalling 
(or collapsing) gas component.

% = = = = = = = = = = = = = = = = = = = = = = = = = = = = = = = = = = = = = = = = = = = = = = = = = =
\section{The Method}
% = = = = = = = = = = = = = = = = = = = = = = = = = = = = = = = = = = = = = = = = = = = = = = = = = =

% ---------------------------------------------------------------------------------------------------
\subsection{Formalism}

The emerging intensity of a beam of radiation in the line of sight after passage through
a scattering atmosphere can be expressed as a convolution of the FRDF and the incoming intensity:

\begin{equation}
  I(x) = \int F(s)\, B(\nu_0)\, ds
,
\end{equation}
where $x = h \nu / k_B T_{CB}$ is the dimensionless frequency,
$B(\nu_0)$ is the incoming intensity (hereafter assumed to be Planckian), and
$F(s)$ is the FRDF, which specifies the probability of scattering from $\nu_0$ to $\nu$ 
as a function of the logarithm of the dimensionless frequency, $\nu/\nu_0$, $s = \ln ( \nu/\nu_0)$, 
where $h$, $\nu$, $k_B$ and $T_{CB}$ are the Planck constant, the frequency, the Boltzmann constant,
and the temperature of the CMBR, $T_{CB} = 2.728 \pm 0.002$ K (\cite{Fixsenet96}).
The FRDF can be expressed as 

\begin{equation}
  F(s) =  e^{-\tau} \delta(s) + w_{sc}\, P(s)
,
\end{equation}
where the first term containing the Dirac delta function, $\delta(s)$, describes
the attenuated incoming radiation by a factor depending on the line of sight optical depth, $\tau$, 
the ``out-scattered'' radiation, and the second term describes the contribution from 
scattering into the beam, which depends on the FRDF of the scattered radiation, $P(s)$,
and a weight, $w_{sc}$, which determines what fraction of the radiation scatters into the beam.
This decomposition is possible because in our approximation the fractional frequency change 
is independent of the frequency (see equation~\ref{e:nu_nu} later). 
The change of the intensity in the line of sight may be expressed as 

\begin{equation} \label{e:DelI}
   \Delta I(x) = \bigl( e^{-\tau} - 1 \bigr) B(x) + \int w_{sc}\, P(s) \,B(\nu_0) ds
.
\end{equation}
For isotropic and homogeneous scattering conditions, so that the scattering
parameters do not depend on where the scattering happens, the weight 

\begin{equation} \label{e:wtau_hom}
   w_{sc} = \bigl( 1 - e^{-\tau} \bigr)
,
\end{equation}
which means that the out-scattered radiation is balanced by the same amount of
in-scattered radiation, and so there would be no net intensity change ($\Delta I(x) = 0$)
if there were no frequency change ($P(s)$ is the Dirac delta function). 
Our task is to calculate $P(s)$ and $w_{sc}$. However, the assumption in 
equation~(\ref{e:wtau_hom}) breaks down where the radiation field is not isotropic
within the cloud, as will be the case in our static and collapsing models, or where 
there is relativistic bulk motion, which introduces anisotropy in the scattering via 
the relativistic beaming effect. 
In the cases which we discuss in the present paper, equation~(\ref{e:wtau_hom}) is an excellent
approximation as we have been able to verify using the results of our Monte Carlo simulations
(see section~\ref{SS:MC_Method}). Significant departures from equation~(\ref{e:wtau_hom}) 
will occur where the scattering optical depth becomes large, or where the gas velocities 
approach the speed of light: the appropriate treatment in these cases is discussed in
a forthcoming paper. Our approximations are adequate for 
clusters of galaxies, thus we are going to assume the validity of equation~(\ref{e:wtau_hom})
in the rest of this paper. 

We derive $P(s)$ using a Monte Carlo method. At low optical depth
($\tau \lesssim 1$), the problem is suitable for Monte Carlo simulation because we do
not have to follow photons through many scatterings in the medium 
(the average number of scatterings being approximately $\tau$).
Although in this limit most of the photons do not scatter, and hence provide no information 
on $P(s)$, this is not a problem since we can use the 
method of forced first scattering (see below).

% ---------------------------------------------------------------------------------------------------
\subsection{Monte Carlo Method}
\label{SS:MC_Method}

We give a short description of the method here, for a more detailed description see 
Molnar (1998). 

We assume an isotropic incoming low temperature radiation field (the CMBR). 
We use forced first scatterings to study the inverse Compton process. 
The photons Compton scatter from an electron population
with a relativistic Maxwellian distribution of momenta in the rest frame of bulk motion. 
We compute scattering probabilities in the rest frame of the electron in the
Thomson limit (\cite{Chandra50}).
This involves coordinate transformations from the observer's
frame to the rest frame of the bulk motion and to the rest frame of the electron. 
We assume time translation invariance and spherical symmetry. 
Time translational invariance is not exact for our model with infall, so that we make a 
snap-shot approximation. 
The error arising from this approximation is less than the light crossing time 
over the infall time ($\approx v_r / c = \beta_r$), which is only 
a few per cent of the infall term for our models.

In most cases we use the inverse method to generate the
desired probability distribution. We use a rejection method when the inverse method leads
to non-invertible functions or is too slow: for a general description of generating probability 
distributions cf. Pozdnyakov et al. (1983); Press et al. (1992). 
The former reference describes an alternative Monte Carlo method to treat inverse
Compton scattering.

In the description that follows we use the word ``photon'' in the singular to refer to 
one Monte Carlo ``photon'', one experiment in our simulation. We use a weight, $w_{in}$,
to express the number of photons this one experiment represents (the weight does not 
have to be an integer). We carried out the simulation in five steps.

% - - - - - - - - - - - - - - - - - - - - - - - - - - - - - - - - - - - - - - - - - - - - - - - - - -
\nop \underline {\it Step 1. The position and direction of incoming photons:}

\nop We assume that the photons arrive uniformly on a unit sphere (the radius of the gas is 
scaled to unity). In a coordinate system which is placed at the point of impact, 
the direction cosine of the incoming photons from the normal, $\mu_{in}$, 
can be sampled as

\begin{equation}
   \mu_{in} = \sqrt{ [RN] }
,
\end{equation}
where we use $[RN]$ to indicate a uniformly distributed random number drawn
each time when it occurs. The azimuthal angle is assumed to be uniformly 
distributed between 0 and $2 \pi$.

% - - - - - - - - - - - - - - - - - - - - - - - - - - - - - - - - - - - - - - - - - - - - - - - - - -
\nop \underline {\it Step 2. distance to the forced first scattering:}

\nop We use forced first scattering, which means that we take the probability of scattering
equal to one on the photons' original line of flight through the gas. 
Using the inversion method, the optical depth at which the incoming photon scatters 
becomes

\begin{equation}   
   \tau_1 = - \ln \bigl( 1- [RN] \cdot (1-e^{-\tau^0_{max}}) \bigr)
,
\end{equation}
where $\tau^0_{max}$ is the maximum optical depth in the line of sight
(the superscript refers to the number of times the photon has already scattered; zero in this case).
Since we are using forced first collisions, we have to account for the fraction of photons 
which are unscattered on their path through the cloud.
The scattered weight may be obtained from equation~(\ref{e:wtau_hom}):
$w_{sc} = \bigl( 1 - e^{-\tau^0_{max}} \bigr) w_{in}$, and stays the same during 
subsequent (unforced) scatterings. 
The weight of the photons passing through the cloud
without scattering is $w_{out} = w_{in} e^{-\tau^0_{max}}$.

% - - - - - - - - - - - - - - - - - - - - - - - - - - - - - - - - - - - - - - - - - - - - - - - - - -
\nop \underline {\it  Step 3. scattering:}
 
\nop At the calculated position of the (forced first) scattering, we use the direction
of the incoming photon and obtain the direction of propagation and frequency of the 
scattered photon. In the case that the gas is moving, we make a Lorentz transformation 
into the rest frame of the moving plasma.
We sample the scattered electron's dimensionless velocity, $\beta_e = v_e/c$, 
from a relativistic Maxwellian distribution

\begin{equation}  
  P(\beta_e)\, d\beta_e = 
                          N_{rel} \, \beta_e^2\, \gamma^5\, e^{-\gamma/\Theta} \, d\beta_e
,
\end{equation}
where the normalization is

\begin{equation} \label{e:N_rel}
  N_{rel} = \Bigl( \Theta \, K_2(1/\Theta) \Bigr)^{-1}
,
\end{equation}
$K_2$ is the second order modified Bessel function of the second kind,
and the dimensionless electron temperature is

\begin{equation} \label{e:theta_e}
   \Theta  = {k_B\, T_e \over m_e c^2}
.
\end{equation}
We used the rejection method to sample $\beta_e$.

The distribution of the direction of electron momenta is simplest in a frame in which 
the photon momentum unit vector points into one of the coordinate axes, z for example. 
In this coordinate system the probability distribution of $\mu_e$, the cosine of the angle 
between the unit vector of the direction of photon propagation (z axis) and electron 
velocity, is

\begin{equation}  
  P(\mu_e, \beta_e) =  (1-\beta_e \mu_e) /2
.
\end{equation}
The inversion method leads to sampling $\mu_e$ as

\begin{equation}  
  \mu_e = {1 \over \beta_e} \Biggl(1 - \sqrt{1-2 \beta_e \biggl(2 [RN]-1 
              - {\beta_e \over 2}\biggr)} \Biggr)
,
\end{equation}
where the sign of the square root was determined so that in the limit of small electron 
velocities we recover the result for an isotropic distribution ($\mu_e = 2 [RN]-1$). 
The angular distribution of the scattered electrons in the plane perpendicular
to the momentum vector of the photon is isotropic 
(at azimuthal angle uniformly distributed between 0 and $2\pi$).
In the rest frame of the electron, the cosine of the polar angle of the incoming photon 
is derived from a Lorentz transformation as

\begin{equation}
  \mu = {-\mu_e + \beta_e \over 1 -\beta_e \mu_e }
,
\end{equation}
where the negative sign in front of $\mu_e$ is appropriate for an {\it incoming} photon.
In the electron's rest frame the scattering probability of a photon coming
in with direction cosine $\mu$ and leaving with direction cosine $\mu^\prime$ 
is given by Chandrasekhar (1950)

\begin{equation} \label{e:chandra}
  f( \mu , \mu^\prime ) = {3 \over 8} 
                                        \Bigl( 1 + \mu^2 \, {\mu^\prime}^2 + 
                   {1 \over 2} \bigl( 1 -  \mu^2 \bigr) \bigl(1 - {\mu^\prime}^2 \bigr) \Bigr)
.
\end{equation} 
$\mu^\prime$ can be sampled using a uniform probability distribution by inversion of

\begin{equation}  
  [RN] = { \int_{-1}^{\mu^\prime} f( \mu , \mu^\prime )  d\mu } = 
             {3 \over 16} \biggl(
             \bigl( \mu^2-{1 \over 3} \bigr) {\mu^\prime}^3 + 
             \bigl( 3 - \mu^2 \bigr) \mu^\prime - {8 \over 9} \biggr)
,
\end{equation}
which leads to a cubic equation for $\mu^\prime$,

\begin{equation}  
   \Bigl( \mu^2- {1 \over 3} \Bigr) {\mu^\prime}^3 + 
   \Bigl( 3 - \mu^2 \Bigr) \mu^\prime - {8 \over 3}\Bigl( 2\, [RN]  + {1 \over 3}\Bigr)
   = 0
.
\end{equation}
This cubic equation has a single real solution with absolute value of $\mu^\prime$ 
less or equal to one. 
We now transfer the direction and frequency of the scattered radiation back to the 
observer's frame.
The dimensionless outgoing frequency of the photon normalized to the incoming 
frequency (expressed with the $s$ parameter) becomes

\begin{equation}  \label{e:nu_nu}
  s = \ln {\nu^\prime \over \nu_0} = \ln \Biggl(
        \gamma_e^2 \Bigl( 1-\beta_e \mu_e \Bigr) \Bigl(1+\beta_e \mu^\prime \Bigr) \Biggr)
.
\end{equation}

% - - - - - - - - - - - - - - - - - - - - - - - - - - - - - - - - - - - - - - - - - - - - - - - - - -
\nop \underline {\it  Step 4. loop over scatterings:}
 
\nop Having the point of scattering, the scattered frequency, and the direction of the
scattered photon, we now sample the optical depth to the next scattering. We do
not use forced scattering, so the optical depth follows from the usual (inverse) method as

\begin{equation}   
  \tau_n =  - \ln  [RN]
,
\end{equation}
where the lower index on $\tau$ refers to the $n$th scattering ($n > 1$), 
and we used $[RN] = 1 - [RN]$, which is correct for uniform probability distributions.
If $\tau$ is less than the maximum optical depth in the direction of the
photon momentum after the previous scattering, $\tau^{n-1}_{max}$, 
the photon is taken to have scattered within the cloud, 
and we calculate the new scattering direction and frequency of the scattered photon
as in step 3. If $\tau_n > \tau^{n-1}_{max}$, the photon escaped (scattered $n-1$ times), 
and we register the impact parameter (the distance between the line of sight and the
center of the spherically symmetric scattering medium), the weight of the photon,
the number of scatterings and $s$, the dimensionless frequency of the escaped photon.

% - - - - - - - - - - - - - - - - - - - - - - - - - - - - - - - - - - - - - - - - - - - - - - - - - -
\nop \underline {\it Step 5. The frequency redistribution function:}

At the end of the simulation we sum the weights in every impact parameter bin
to check our assumption of homogeneous scattering (equation~\ref{e:wtau_hom}) and we determine
how the average number of scatterings depend on the impact parameter. 
It turns out that, in our case of low optical depth and very mild infall velocities,
the average number of scatterings has no noticeable dependence on the impact parameter, 
and that equation~(\ref{e:wtau_hom}) is satisfied at each impact parameter within the accuracy of our 
Monte Carlo simulations, which is less than 0.1 \%.  
The scattered FRDF depends only on the number of scatterings, thus we can sum all the photons to 
determine an average scattered FRDF, which can be used as an excellent approximation to the FRDF
corresponding to an arbitrary line of sight.

Therefore the discrete probability distribution of the scattered FRDF can be derived
by binning the frequencies of all the out-coming photons as

\begin{equation}
  P(s_k) =  N_s \, {\cal N}_k
.
\end{equation}
${\cal N}_k$ is the number of photons in the k$\rm ^{th}$ bin, 
for which ${s_k - \Delta s/2} \le s < {s_k + \Delta s/2}$, where  
$s_k$ is the center of the k$^{th}$ bin, $\Delta s$ is the width of the bin, and
$N_s$ is the normalization, $N_s = 1/ (\Delta s N_{MC})$.
$N_{MC}$ is the number of Monte Carlo photons.
This $P(s_k)$ is our sampled approximation to $P(s)$, and we then 
fit an exponential of polynomials to $P(s_k)$ to get a convenient expression for the FRDF.
Our fit gives an approximation accurate to better than half a percent, 
except in the (small) extended tails, where $P(s)$ is under-represented.
However, these regions lie 3 orders of magnitude below the peak, and the error arising 
from the fit is negligible for our applications.

% ---------------------------------------------------------------------------------------------------
\subsection{Testing the code}

% - - - - - - - - - - - - - - - - - - - - - - - - - - - - - - - - - - - - - - - - - - - - - - - - - -
\subsubsection{Single-scattering approximation}

We tested our code by comparing our single scattering Monte Carlo results for the FRDF 
($P_1^{MC}$) to those derived from Rephaeli (1995a).
Rephaeli's single-scattering approximation can be written as

\begin{equation} \label{e:P1s}
  P_1(s) = {3 \over 32\,N_{rel} }\,
          \int_{\beta_0}^1 \gamma\, e^{-\gamma/ \Theta} \beta_e^{-4}
          \bigl( f_1 + f_2 + f_3 \bigr) \,d\beta_e
,
\end{equation}
where the lower limit $\beta_0$ is the minimum $\beta_e$ needed to get the frequency shift $s$, 

\begin{equation}    
   \beta_0 =  {e^{|s|} - 1 \over e^{|s|} + 1} 
.
\end{equation}
The functions $f_1$, $f_2$ and $f_3$ are

\begin{eqnarray} \label{e:P1s2}
  f_1 & =  &    e^{3s} \beta_e^3 \bigl( \mu_2^3 - \mu_2 - \mu_1^3 + \mu_1 \bigr)   \cr
  f_2 & =  &    \biggl( \beta_e^2-3\bigl( 4e^s+1\bigr)+3 \,e^s (x_2+x_1)+
                   { 2\beta_e^2-3(1+\beta_e^4) \over x_2\,x_1} \biggr) e^s\,(x_2-x_1) \cr
  f_3 & =  &    2e^s\, \bigl(e^s + 1\bigr) \bigl(3 - \beta_e^2\bigr) \ln{x_2 \over x_1}  
,
\end{eqnarray}
where $x_1 = 1-\beta_e\,\mu_1$, $x_2 = 1-\beta_e\,\mu_2$, 
$\mu_1$ and $\mu_2$ are defined by 

\begin{eqnarray}                    \label{e:mu1mu2}
   \mu_1 & =  &   \cases{
                   -1                                & $s \le 0$ \cr
                   {1 - e^{-s}(1+\beta_e) \over \beta_e} & $s \ge 0$ \cr
            } \cr
   \mu_2 & =  &   \cases{
                   {1 - e^{-s}(1-\beta_e) \over \beta_e} & $s \le 0$ \cr
                    1                                & $s \ge 0$ \cr
            }
,
\end{eqnarray}
and we used equation~(\ref{e:nu_nu}) to eliminate $\mu\prime$.
This result (equation~\ref{e:P1s}) agrees with Fargion et al. (1996). 
These expressions can be integrated numerically, except when $s = 0$.
In that case $\beta_0 = 0$, and direct numerical integration is not 
possible because of the diverging $\beta_e^{-4}$ term. 
For small $\beta_e$ ($\beta_e < b = 0.1$ for example) we can expand the logarithm, 
and use this expansion as a good approximation. 
For $s = 0$, equation~(\ref{e:P1s}) becomes

\begin{equation}   \label{e:P1s_0}
  P_1(s=0) = {3 \over 8\,N_{rel} }\, 
                      \Biggl( \int_0^b f_0  \,d\beta_e + \int_b^1 f_0  \,d\beta_e \Biggr)
,
\end{equation}
where the integrand is 

\begin{equation}   
  f_0 = \gamma\, e^{-\gamma/ \Theta} \beta_e^{-4} \Biggl( 2\gamma^2 \, \beta_e^5 + 
            6\beta_e+ (3-\beta_e^2) \ln \Bigl({1-\beta_e\over 1+\beta_e}\Bigr) \Biggr)
.
\end{equation}
A Maclaurin expansion of the logarithm to order $\beta_e^4$ gives an adequate 
approximation 

\begin{equation}   
  \int_0^b f_0 \, d\beta_e \approx 
                             \int_0^b \gamma\, e^{-\gamma/ \Theta} \beta_e\,
                                      \biggl( \gamma^2 + {1\over 3} - {3\over 5}
                                              +   \Bigl( {1\over 5} - {3\over 7} \Bigr)  \beta_e^2
                                          + \Bigl( {1\over 7} - {1\over 3} \Bigr)  \beta_e^4 \biggr)
                             \, d\beta_e
\end{equation}
with no remaining divergent terms for the first integral in equation~(\ref{e:P1s_0}).

We derive $P_1^{MC}(s)$ from our Monte Carlo simulation by using the results of only 
the first (forced) scatterings.
% Figure~\ref{f:Ps_MCR1} Figure~\ref{f:Ps_MCR3} 
Figures 1a and 1b show our Monte Carlo results, $P_1^{MC}$, superimposed on 
$P_1$ from equations~(\ref{e:P1s}) and (\ref{e:P1s2}) for dimensionless temperatures 
$\Theta = 0.03$ and 0.3. 
The agreement is excellent, confirming that our Monte Carlo code is successfully 
reproducing $P_1(s)$.

% - - - - - - - - - - - - - - - - - - - - - - - - - - - - - - - - - - - - - - - - - - - - - - - - - -
\subsubsection{Testing the Numerical Integral for the Intensity Change}

We derive the intensity change from the FRDF using a numerical integral (equation~\ref{e:DelI}).
The Kompaneets approximation leads to the following FRDF:

\begin{equation}  \label{E:P_K}
     F_K = { 1 \over \sqrt{ 4 \pi y} } \exp  \Bigl(  - {(s - 3y)^2  \over4y  }  \Bigr) 
,
\end{equation}
where the Compton $y$ parameter is

\begin{equation}   \label{E:Compton_y}
   y = \int n_e \sigma_T \,\Theta d \ell 
,
\end{equation}
where $n_e$ is the electron number density as a function of length in the line of sight measured
by $\ell$ and $\Theta$ is the dimensionless temperature (for a discussion see for example 
\cite{Birkinshaw99}; \cite{Molnar98}).
In order to check our numerical method, we used the Kompaneets FRDF (equation~\ref{E:P_K})
in the numerical integral in equation~(\ref{e:DelI}), and compared the resulting intensity
change to that of obtained by the analytic solution for the Kompaneets approximation

\begin{equation}  
    \Delta I_K = y\, {\imath_0 x^4 e^x \over (e^x - 1 )^2 } 
                     \biggl(  x { e^x + 1 \over e^x - 1 } - 4 \biggr)
, 
\end{equation}
where $\imath_0 = 2 (k_B T_{CB})^3 / (hc)^2$.
We concluded that our numerical method is accurate better than 0.1\%.

% - - - - - - - - - - - - - - - - - - - - - - - - - - - - - - - - - - - - - - - - - - - - - - - - - -
\subsubsection{Relativistic Corrections to Kompaneets equation}

We also compare our results to those from the extended Kompaneets equation up to the
5th order in $\Theta$ (\cite{Itohet98}). Itoh et al. expressed the intensity change as

\begin{equation}   
  \Delta I = { \Delta n \over n} { \imath_0 x^3 \over e^x - 1 }
,
\end{equation}
and provided expressions for ${ \Delta n \over n}$ (note, that their $y$ parameter is 
actually $\tau$, the optical depth).
On Figure~\ref{f:Deli_noz} we plot $\Delta I / \tau$ from the Kompaneets approximation, 
from Itoh et al.'s expansion, and for our single scattering Monte Carlo 
results. From the figure we conclude that our single scattering Monte Carlo result
agrees with that of Itoh et al. at low temperatures, $\Theta < 0.03$ ($T_e < 15$ keV).
Deviations from the  Itoh et al.'s result are already appearing at $\Theta = 0.03$, and 
become more pronounced at higher temperatures and high frequencies, as we would expect.

We conclude that our simulation method passes these two tests, and can now be used to
calculate the effects of multiple scattering and bulk velocity on the SZ effect.

% = = = = = = = = = = = = = = = = = = = = = = = = = = = = = = = = = = = = = = = = = = = = = = = = = =
\section{Results}
% = = = = = = = = = = = = = = = = = = = = = = = = = = = = = = = = = = = = = = = = = = = = = = = = = =

We performed a number of simulations for uniform density spherical models which are
either static or have radial infalls with constant gas velocity at all radii. 
These models cover a range of $\tau_0$, the optical depth of zero impact parameter,
electron temperature, $T_e$, and gas infall speed, $\beta_r$. 
Simulations with monochromatic incoming radiation were used to determine the scattered FRDF.

We verified via our simulations that for our low optical depth static models and for our 
models with low optical depth and very mild infall velocities, a parameter space adequate for
clusters of galaxies, the dependence of the scattered FRDFs on the impact parameter is 
negligible, and that equation~(\ref{e:wtau_hom}) provides a very good approximation to the weight
of the scattered radiation. 
Although we determined an averaged scattered FRDF from all scattered photons regardless of their 
impact parameter, in our case, the determined scattered FRDF can be used at any impact parameter,
since the dependence of the average number of scatterings on the impact parameter is negligible 
(see section~\ref{SS:MC_Method}, Step 5). 

In Figure~\ref{f:Ps_sta_a} we show the FRDFs of scattered photons emerging from our spherical 
static models with maximum optical depth $\tau_0 = 0.05$ and seven different temperatures. 
We used all photons to derive the scattered FRDFs. At higher temperatures more photons scatter
into higher energies, thus the FRDFs are broader, and have lower peaks (since they
are normalized to unity). 
We show the effect of finite optical depth in Figure~\ref{f:Ps_sta_b}: higher optical depth leads 
to more scatterings, and therefore more photons scattered to higher energies.
Even for an optical depth as large as $\tau_0 = 0.1$ the change in the function is relatively 
small. In Figure~\ref{Ps_col_b} we show the effect on the FRDF of gas infall. 
Larger infall velocities cause more up-scattering of the photons, 
and hence more spreading of the FRDF, but the most obvious change is that the 
sharp peak at $s = 0$ is smoothed out by the motion of the plasma. 
At lower temperatures bulk motion causes larger departures from the static FRDF since the 
infall speed is larger relative to the electron thermal velocity. 

We used these results for the scattered FRDF to calculate intensity change using 
equations~(\ref{e:DelI}) and (\ref{e:wtau_hom}).  
We evaluated the emergent intensity change at zero impact parameter  
(i.e. through the center of the gas sphere), where $\tau = \tau_0$.
In Figure~\ref{f:Deli_stat} we show the intensity change $\Delta I$ for a static plasma for 
two optical depths and five temperatures. 
Non-zero optical depth causes only slight changes in the emerging radiation.
Figure~\ref{f:Deli_coll} shows the intensity change for a plasma with infall for two
infall velocities and three plausible cluster temperatures. Only small changes in the spectrum 
are apparent, even with such large velocities.

One measure of the spectral deformation that has been used to quantify relative correction, 
and which is of use in determining the frequency at which to search for the KSZ effect, is
the cross-over frequency.
Challinor and Lasenby (1998a) and Birkinshaw (1998) suggested a linear expression for how the 
frequency changes with temperature, as

\begin{equation}   \label{e:X0_LIN}
  X^{lin}_0 = 3.830 ( 1 + 1.13 \, \Theta)
,
\end{equation}
while \cite{Itohet98} suggested a quadratic approximation

\begin{equation}  \label{e:X0_Q}
  X^{q}_0 = 3.830 ( 1 + 1.1674 \, \Theta - 0.8533  \, \Theta^2)
\end{equation}
for most cluster temperatures.
In Figure~\ref{f:x0_MC_NOZ} we compare our Monte Carlo results with these and other 
expressions that include relativistic corrections.
Our Monte Carlo results for single scattering are close to those obtained by 
numerical integration of the collision integral (\cite{Nozawaet98}).
Relativistic corrections of third and fifth order (\cite{Challinor98a}; \cite{Nozawaet98}), or the 
linear approximation (\cite{Challinor98a}; \cite{Birkinshaw99}) are of varying accuracy
in describing the curve: the linear and third order expressions give the best results, 
but extending the series to the fifth order is much poorer.
This is a consequence of the asymptotic nature of the series, as emphasized by 
Challinor and Lasenby (1998a).

Figures \ref{f:x0_MC_STAT} and \ref{f:x0_MC_COLL} show the cross-over
frequency as a function of dimensionless temperature $\Theta$ for finite optical depth 
and infall velocity. Including a finite optical depth causes only a small change in the
$X_0(\Theta)$ curve, and this change does not depend much on temperature. 
Based on our Monte Carlo simulations, we suggest the following approximation
for the cross-over frequency for single scatterings in static spherical plasma for 
dimensionless temperature $\Theta \lesssim 0.3$:

\begin{equation}   
  X_0^s(\Theta) = 3.827\,( 1 + 1.2038\, \Theta -1.2567 \,\Theta^2 
                                                + 0.9098 \,\Theta^3)
,
\end{equation}
which fits better than 0.01 \% 
in this range with a shift from the optical depth dependence

\begin{equation}   
 \Delta X_0^\tau(\tau_0) =  \tau_0 \bigl( 0.35 - 0.04416\, \Theta^{-0.5} \bigr)
,
\end{equation}
which fits better than 0.005 \%  
for  $0.05 \le \Theta \le 0.3$, and about 0.1 \% 
for lower temperatures, and $\tau_0 < 0.5$.

Figure~(\ref{f:x0_MC_COLL}) shows results for our spherical models with gas infall. 
As for non-zero optical depth, additional energy transfers occur because of motion of the gas. 
As we would expect, at low temperatures the contribution to the electron velocity from 
bulk motion is comparable to that from thermal motion, and an enhanced frequency
shift results, while at high temperatures this contribution becomes negligible.
Based on our Monte Carlo simulations, we suggest the following approximation
for the cross-over frequency shift in plasma with infall for $0.01 \le \Theta \le 0.3$:

\begin{equation}   \label{e:x0_beta}
    \Delta X_0^\beta (\Theta, \beta_r) = 0.224 \, \beta_r^2\, \Theta^{-0.7}
.
\end{equation}
This formula fits the cross over frequency $\nu_0$ better than about half a percent.

At small optical depth and bulk velocity we can assume that the shifts simply add, 
so that the final expression for the cross-over frequency becomes

\begin{equation}   
   X_0(\Theta, \tau_0, \beta_r) = X_0^s(\Theta) + 
                                     \Delta X_0^\tau(\tau_0)  + \Delta X_0^\beta(\Theta, \beta_r)
.
\end{equation}

% CLUSTERS  CLGXs

In Figure~\ref{f:x0_GLGX} we show the cross-over frequency as a function of 
temperature and optical depth in the parameter range ($\Theta  \lesssim 0.04$, 
$\tau_0 \lesssim 0.05$) important for clusters of galaxies.
From this figure we may come to the conclusion that, for clusters of galaxies, 
the optical depth effect on the cross-over frequency is more important than 
non-linear terms in the expansion in $\Theta$.
We provide more accurate fitting formulae for this range. From our models we find

\begin{equation}   \label{e:x0_GL}
 X_0^{cl}(\Theta, \tau_0, \beta_r) = 
          3.829\,( 1 + 1.1624\, \Theta - 0.68948 \,\Theta^2) + 0.14\, \tau_0
        + \Delta X_0^\beta(\Theta, \beta_r)
,
\end{equation}
which fits better than 0.03 \%
for static models in this range of parameters. 
The optical depth independent first term is similar
to the approximation provided by Itoh et al. (1998), which was
obtained by numerically integrating the collision integral.

The relativistic corrections to the kinematic SZ effect 
(cluster radial bulk velocity, $v_{rad}$)
have also been found to be important (\cite{Nozawaet98}, \cite{SazonovSun98b})
The shift was found to be

\begin{equation}   
   \Delta X_0^{kin}(\Theta, v_{rad}) = 300\,{v_{rad} \over c}
           \biggl[ { a_1 \over \Theta - \Theta_{min} } + { a_2 \over (\Theta - \Theta_{min})^2}
                             + a_3 + a_4 \, \Theta + a_5 \, \Theta^2
              \biggr]
,
\end{equation}
where $v_{rad}$ is the radial velocity, 
$\Theta_{min} = 1.654 \times 10^{-3}$, $a_1 = 3.857 \times 10^{-3}$, 
$a_2 = -4.631 \times 10^{-6}$, $a_3 = 1.370 \times 10^{-2}$, $a_4 = 1.014 \times 10^{-2}$, 
and $a_5 =  0.01$ (\cite{Nozawaet98}), which agrees well with Sazonov and Sunyaev's (1998b) 
result (expressed by a different fitting function).
In Figure~\ref{f:x0_GLGX}, long dashed lines represent the results of this kinematic effect
of cluster radial velocity of $\pm$~100~\Kms $\,$
($X_0^{cl}(\Theta,  0) + \Delta X_0^{kin}(\Theta, v_{rad})$)
Comparing our results for the shift in the cross-over frequency to results from
relativistic kinematic effect, we conclude that
a shift caused by an optical depth of 0.01 is equivalent to a shift caused by a radial velocity
of about $v_{rad} = -10$ \Kms.

We estimate the amplitudes of these effects on the hot cluster, Abell 2163, which was 
discussed by Holzapfel et al. (1997a). The maximum optical depth of the cluster
is $\tau_0 = 0.01$, the temperature of the intracluster gas is close to $\Theta = 0.03$.
Using our results for the static effect with the given temperature and maximum optical 
depth we get about 100 MHz shift to higher frequencies relative to the linear expression
of Challinor and Lasenby (1998a). An infall velocity of $\beta_r = 0.01$ causes about an additional
15 MHz shift to higher frequencies relative to our result for the static model.
These shifts are small relative to the 20 GHz band width of the instrument of 
Holzapfel et al. (1997a) and the error in $H_0$ and cluster radial peculiar velocity from ignoring 
their presence would be about 10 \% (if the SZ effect is measured at $x \gtrsim 5$) 
and 15 \Kms. By comparison, the component of primordial anisotropy in this 
scale corresponds to adding a velocity noise about $\pm \, 200$ \Kms.

% = = = = = = = = = = = = = = = = = = = = = = = = = = = = = = = = = = = = = = = = = = = = = = = = = =
\section{Conclusions}
% = = = = = = = = = = = = = = = = = = = = = = = = = = = = = = = = = = = = = = = = = = = = = = = = = =

We investigated the effect of finite optical depth and bulk motion on inverse
Compton scatterings in spherically symmetric uniform density mildly
relativistic plasma. We assumed isotropic incoming radiation (CMBR),
a relativistic Maxwellian distribution for the electron momenta, and
scatterings in the Thomson limit. We demonstrated
the usefulness of our Monte Carlo method for solving the radiative 
transfer problem, and calculated the static and kinematic SZ effects with different 
optical depth and gas infall velocities.

The solution of the extended Kompaneets equation (with corrections up to the fifth order) 
is equivalent to a single-scattering approximation, and significant deviations from it 
occur for hot clusters and at high frequencies. These deviations may be as large as 5 \% 
of the intensity change and neglecting them could cause about a 10 \% 
error in the Hubble constant.
A finite optical depth causes further small changes in the SZ effect: 
these changes may exceed the relativistic correction terms. 
For typical cluster temperatures, an accurate expression for the cross-over frequency
as a function of temperature, optical depth, and bulk motion is (\ref{e:x0_GL}).

As it can seen from Figure~\ref{f:x0_GLGX}, the cross-over frequency is 
sensitive to the cluster radial velocity, and less sensitive to the finite optical depth. 
Measurements of the cross-over frequency can, in principle, be used to determine the 
radial velocity of the cluster (e.g., as in \cite{Holzapfelet97b}), with small extra corrections
for optical depth and possible gas motion inside the cluster.
However, the relatively strong variation of $X_0$ with $\Theta$, compared to $\tau$
or $\beta_r$, suggests that the largest uncertainty will arise from the assumption of 
cluster isothermality, even if effects of confusion from primordial (and secondary) 
CMBR fluctuations can be excluded.

Finally we note that our method can be extended to any geometry, density distribution and 
complicated bulk motion as desired, and may be used to study the SZ effect in 
high temperature plasmas with or without bulk motion.

\acknowledgments

SMM is grateful to Bristol University for a full scholarship, where most of this work was done. 
This work was finished while SMM held a National Research Council-NASA/GSFC Research Associateship. 
We thank to our referee, Dr. Challinor, for suggestions which helped us to clarify the
presentation of the results.

% % % % % % % % % % % % % % % % % % % % % % % % % % % % % % % % % % % % % % % % % % % % % % % % % % %
% 
%                                   B I B L I O G R A F Y
% 
% % % % % % % % % % % % % % % % % % % % % % % % % % % % % % % % % % % % % % % % % % % % % % % % % % %

% % % % % % % % % % % % % % % % % % % % % % % % % % % % % % % % % % % % % % % % % % % % % % % % % % %
% 
%                                      F I G U R E S OUT
% 
% % % % % % % % % % % % % % % % % % % % % % % % % % % % % % % % % % % % % % % % % % % % % % % % % % %

\clearpage

%  FIGURE 1a  FIGURE 1b
\begin{figure} 
\epsfig{file=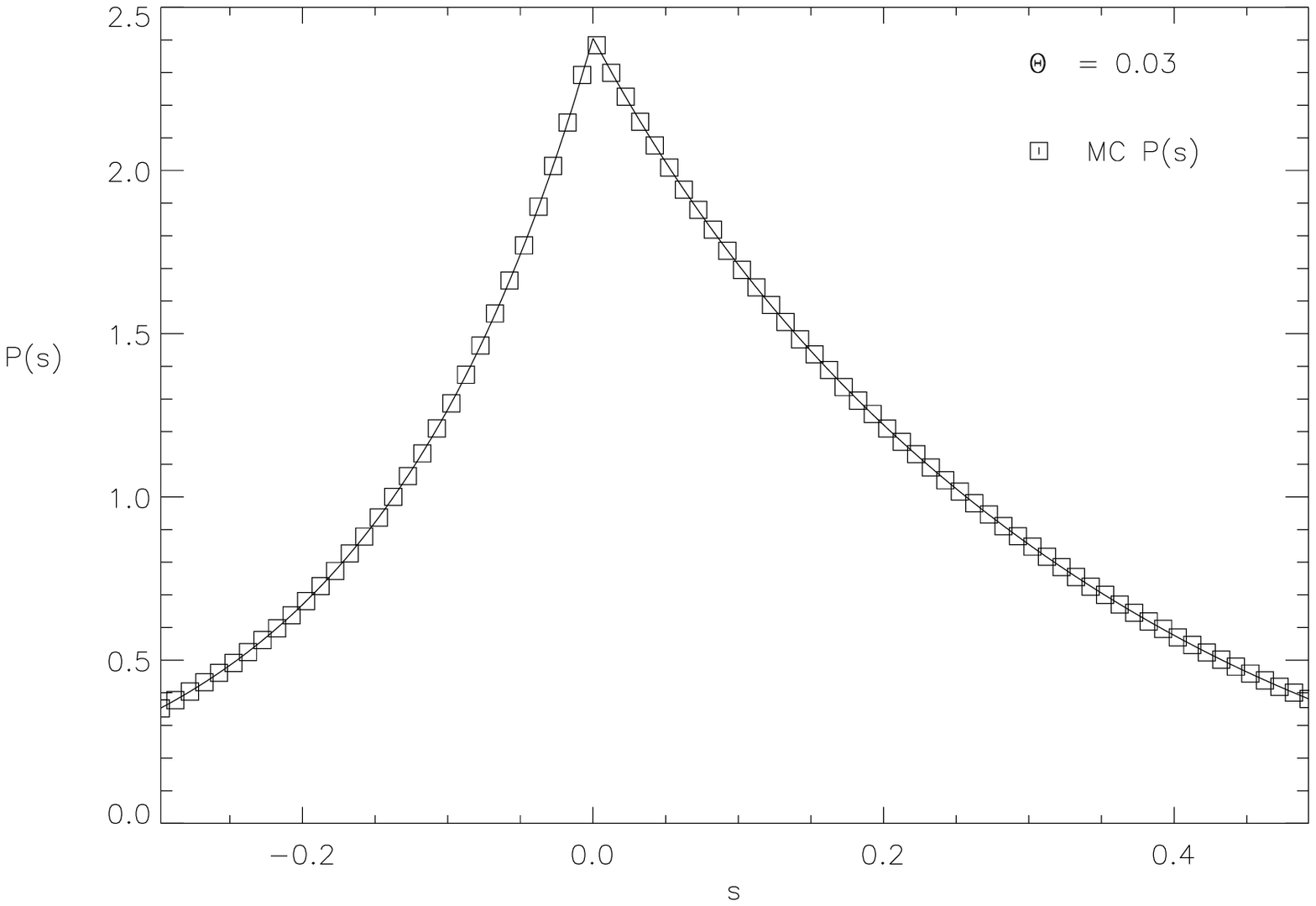,height=80mm}
\epsfig{file=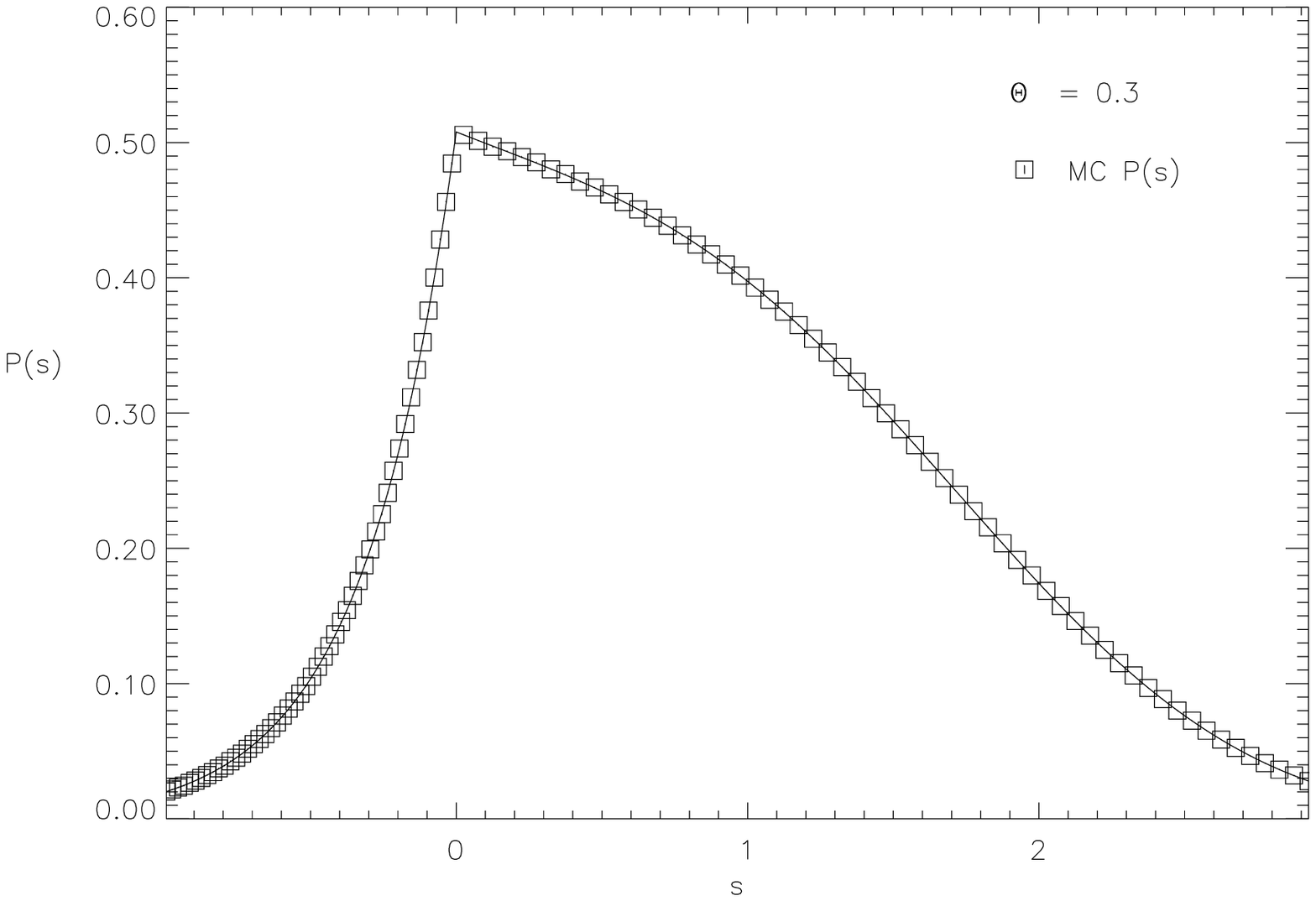,height=80mm}
\caption{\label{f:Ps_MCR1}
   The frequency redistribution function, $P(s)$, for a static, single scattering case  
   at dimensionless temperatures $\Theta = 0.03$ (a), or $\Theta = 0.3$ (b).
   The solid line shows the result from Rephaeli 1995b)'s semi-analytic method, 
   the boxes with vertical error bars (too small to be visible) show results from 
   our Monte Carlo simulations. 
}
\end{figure}

%  FIGURE 2
\begin{figure} 
\epsfig{file=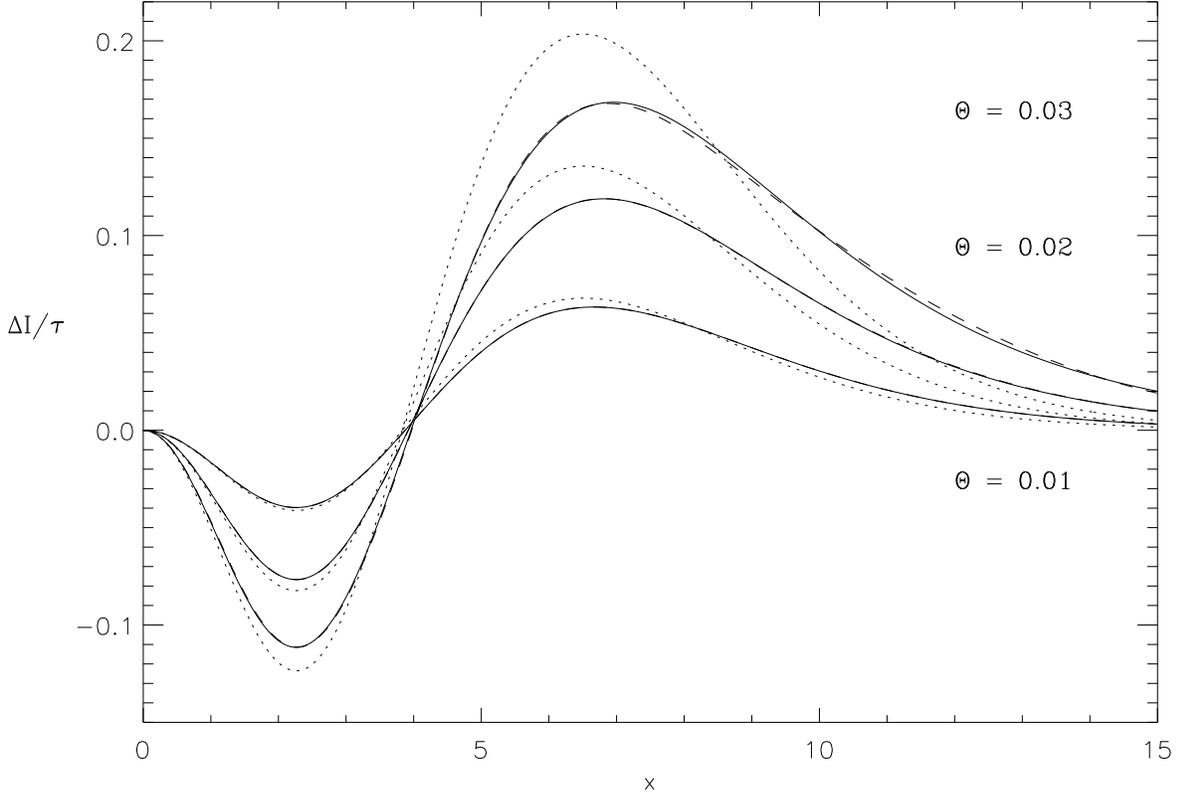,height=115mm}
\caption{\label{f:Deli_noz}
   The intensity change $\Delta I / \tau$ (in units of $\imath_0 = 2 (k_B T_{CB})^3 / (hc)^2$) 
   as a function of dimensionless frequency $x = h \nu / (k_B T_{CB})$ for  
   dimensionless temperatures $\Theta = 0.01$, 0.02, and 0.03 in static spherically 
   symmetric models. The solid, dashed and dotted lines are results from  
   Monte Carlo method (single scattering), from relativistic corrections to the  
   Kompaneets approximation (Itoh et al. 1997)  and the  
   Kompaneets approximation (Kompaneets 1957) respectively. Note that 
   at  low temperatures, $\Theta = 0.01$ and 0.02, 
   the single scattering Monte Carlo method 
   and the Kompaneets approximation with relativistic corrections give very similar 
   results (the solid and dashed lines overlap). 
}
\end{figure}

%  FIGURE 3
\begin{figure} 
\epsfig{file=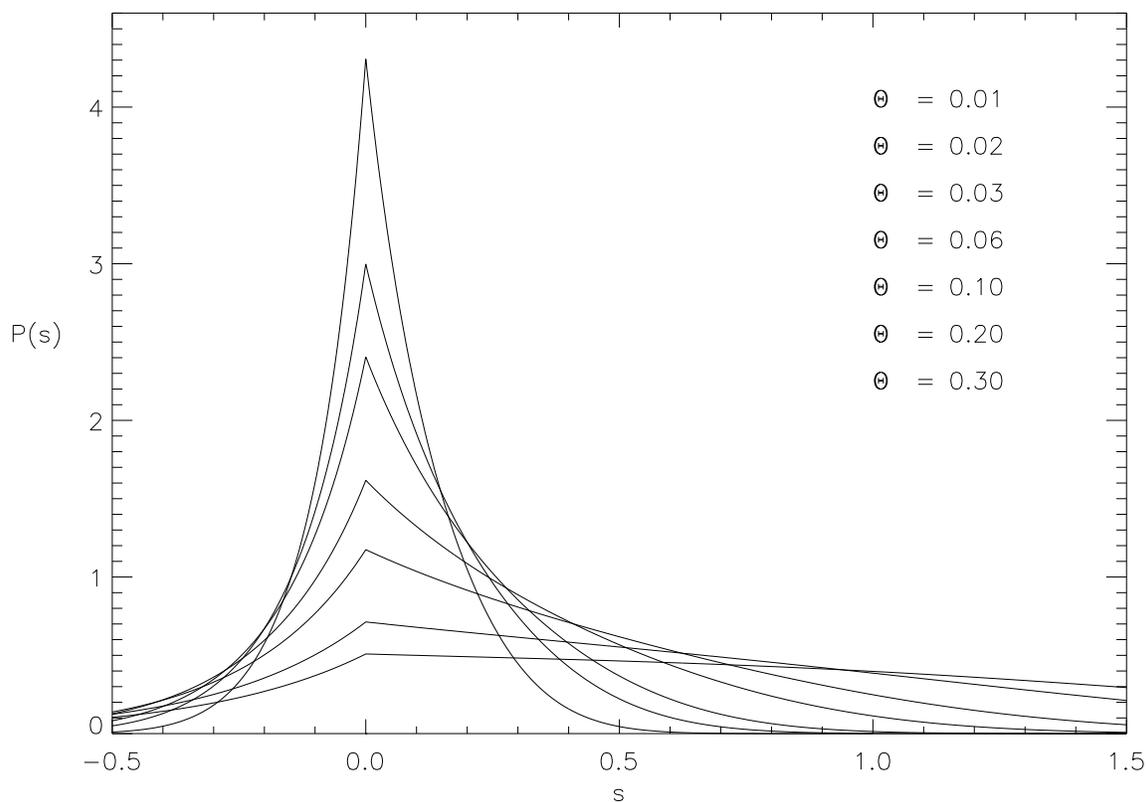,height=115mm}
\caption{\label{f:Ps_sta_a}
   Scattered frequency redistribution functions, $P(s)$-s, of scattered photons emerging from
   our spherical static models with maximum optical depth $\tau_0 = 0.05$ 
   and seven different temperatures. 
   We used all emerging photons to determine $P(s)$ regardless of their impact parameter.
   (see text for details).
   The higher the temperature, the lower the peak and wider the function due to 
   larger energy transfers from the hot electrons to the photons.
}
\end{figure}

%  FIGURE 4
\begin{figure} 
\epsfig{file=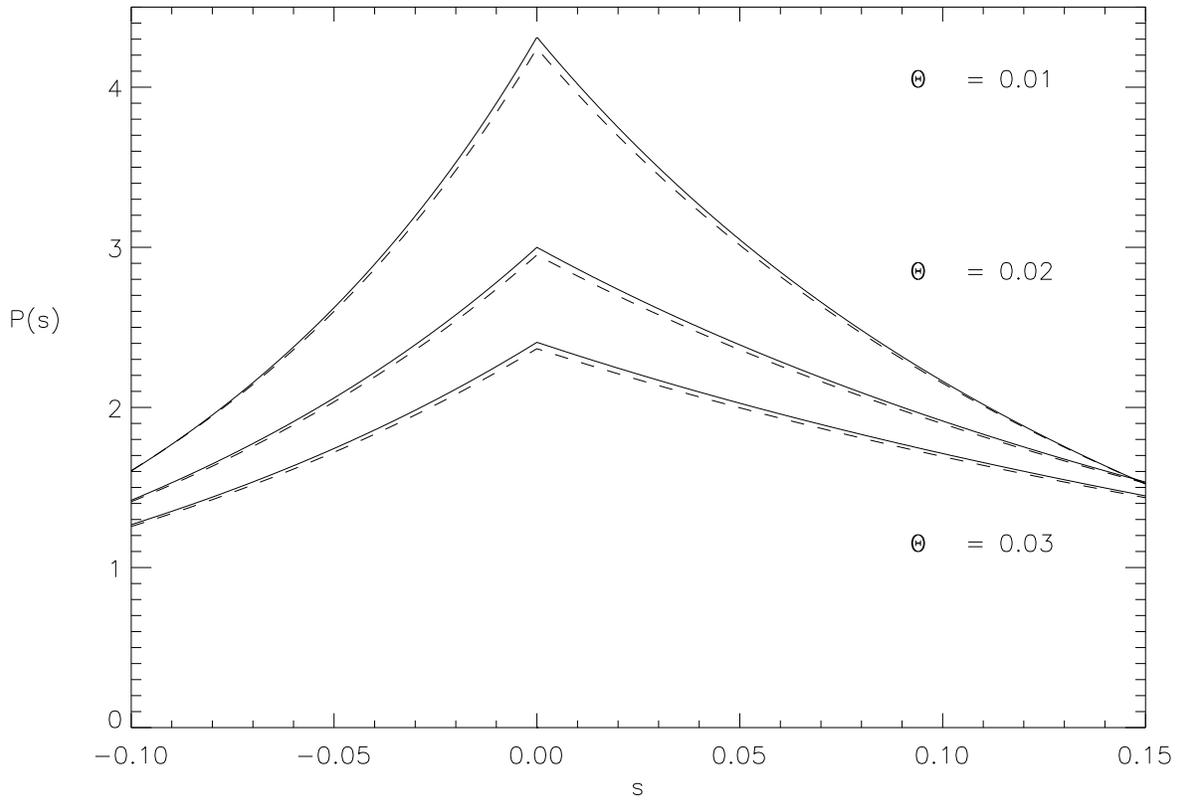,height=115mm}
\caption{\label{f:Ps_sta_b}
   The inner part of the scattered frequency redistribution function, $P(s)$, 
   for static clusters with dimensionless temperatures $\Theta = 0.01$, 0.02, and 0.03. 
   The solid and dashed lines show the extremes of 
   single scattering and scattering with maximum optical 
   depth $\tau_0 = 0.1$. More scatterings lead to more energetic photons, 
   and hence more scattering from the line center to the high energy tail.
} 
\end{figure} 

%  FIGURE 5
\begin{figure} 
\epsfig{file=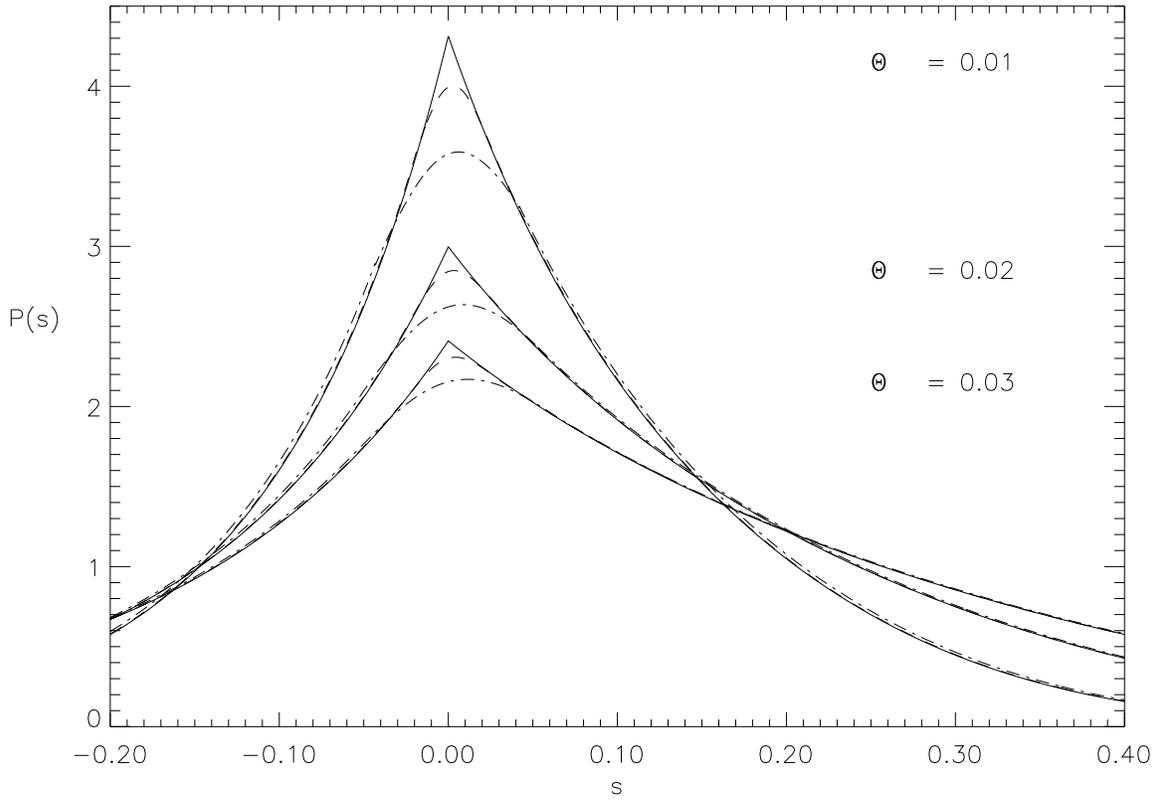,height=115mm}
\caption{\label{Ps_col_b}
   The single scattering frequency redistribution function, $P(s)$, for infalling plasma 
   with dimensionless temperatures $\Theta = 0.01$, 0.02 and 0.03. 
   The solid, dashed and dashed dot lines show results for models with infall velocities
   $\beta_r = 0$ (static), 0.03, and 0.05.
} 
\end{figure} 

%  FIGURE 6
\begin{figure} 
\epsfig{file=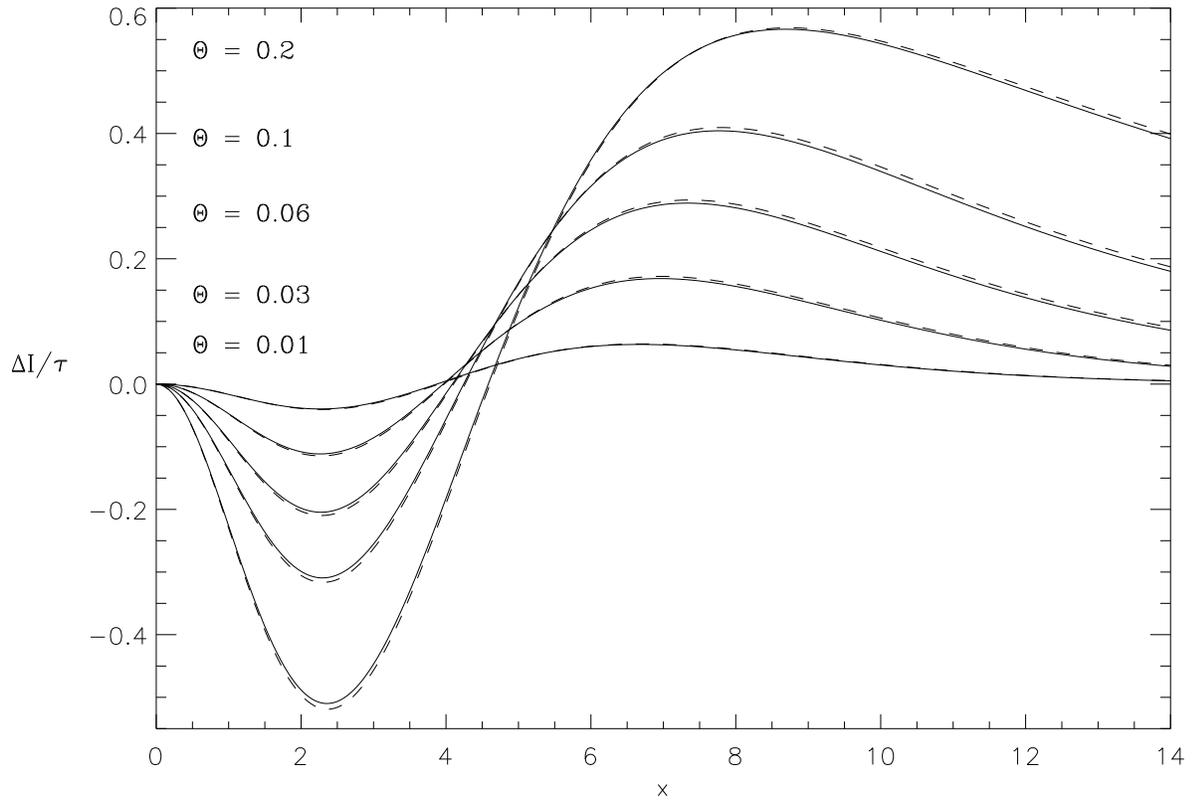,height=115mm}
\caption{\label{f:Deli_stat}
   The intensity change $\Delta I / \tau$ (in units of $\imath_0 = 2 (k_B T_{CB})^3 / (hc)^2)$)
   as a function of dimensionless frequency $x = h \nu / (k_B T_{CB})$ for five 
   dimensionless temperatures in static spherically symmetric models, 
   for single scattering (solid lines) and $\tau_0 = 0.1$
   (dashed lines).
}
\end{figure} 

%  FIGURE 7
\begin{figure} 
\epsfig{file=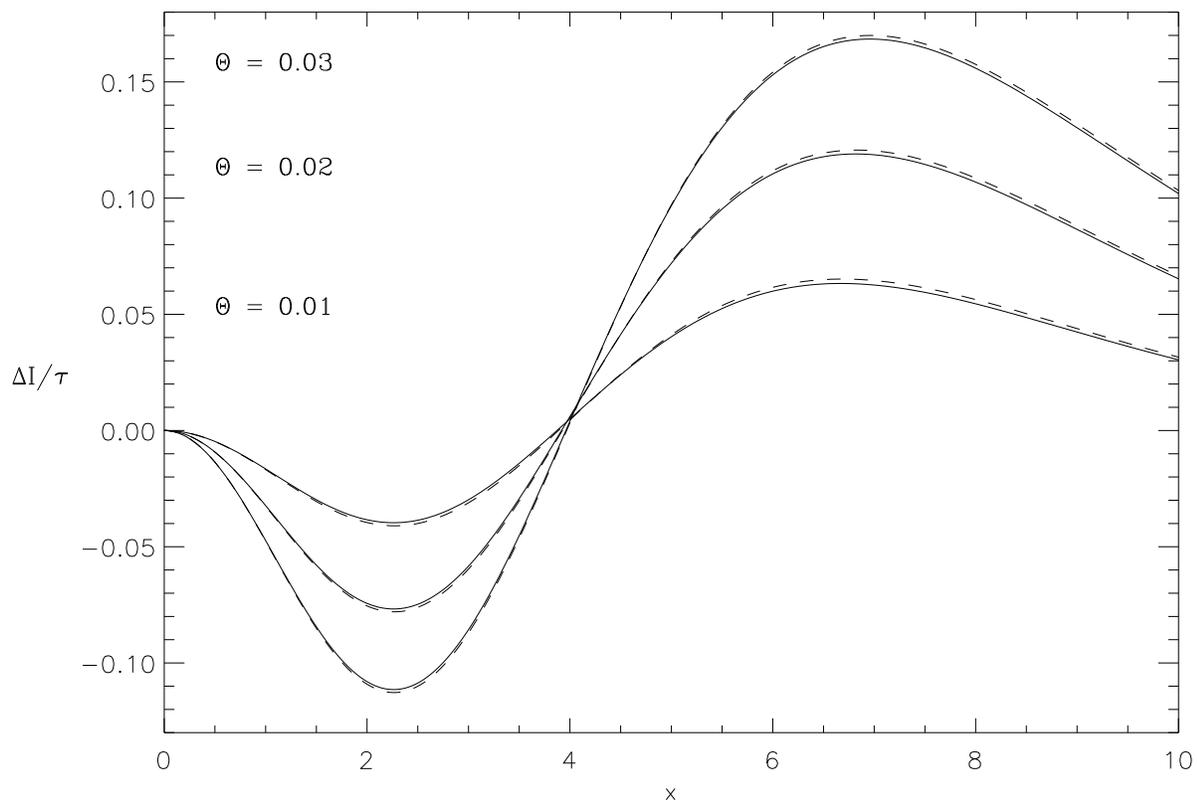,height=115mm}
\caption{\label{f:Deli_coll}
   The intensity change $\Delta I / \tau$ (in units of $\imath_0 = 2 (k_B T_{CB})^3 / (hc)^2)$)
   as a function of dimensionless frequency $x = h \nu / (k_B T_{CB})$ for three 
   dimensionless temperatures in spherically symmetric models with infall. Monte Carlo
   results are plotted using single scattering approximation for static gas (solid line)
   and gas with bulk motion ($\beta_r = 0.05$, dashed line). 
}
\end{figure}      

%  FIGURE 8
\begin{figure} 
\epsfig{file=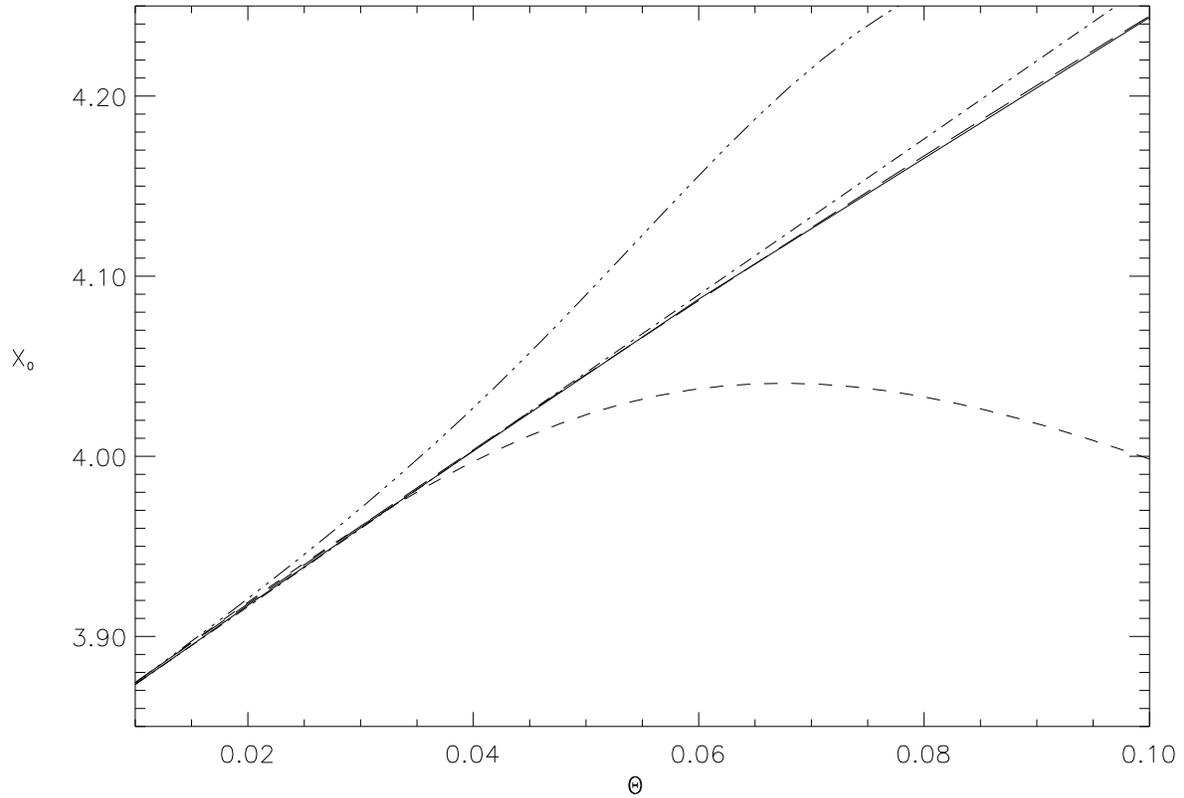,height=115mm}
\caption{\label{f:x0_MC_NOZ}
     The cross-over frequency as a function of dimensionless temperature, $\Theta$. 
     The solid and long dashed lines show our Monte Carlo results for single scattering, 
     and the results of a numerical evaluation of the collision integral (Nozawa et al 1998).
     The dashed dot line represents a linear approximation 
     (Challinor and Lasenby 1998a; Birkinshaw 1998).
     The short dashed and dashed dot dot dot lines use relativistic corrections
     of third and fifth order in $\Theta$.
}
\end{figure} 

%  FIGURE 9
\begin{figure} 
\epsfig{file=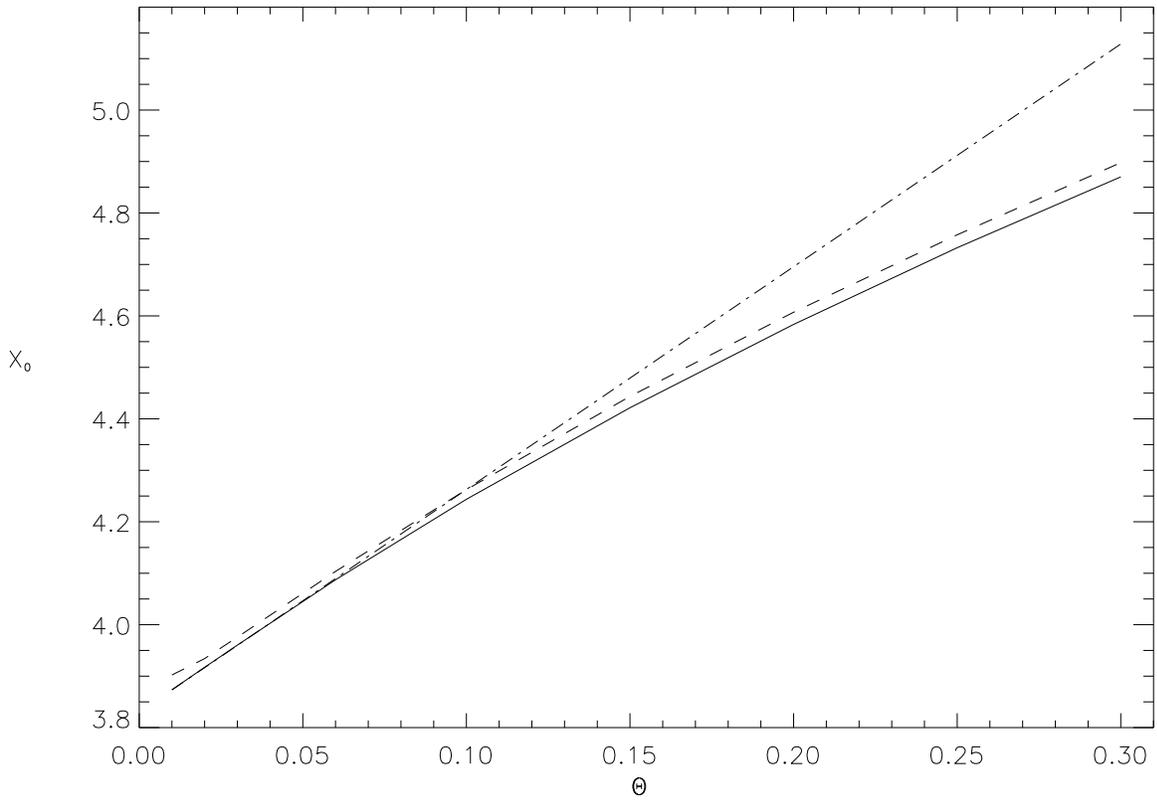,height=115mm}
\caption{\label{f:x0_MC_STAT}
     The cross-over frequency as a function of dimensionless temperature,  
     $\Theta$ for static, spherically symmetric plasma.  
     The solid and dashed lines are our results for single scattering and $\tau_0 = 0.1$,
     respectively. 
     The dashed dot line shows a linear approximation 
     (Challinor and Lasenby 1998a; Birkinshaw 1998).
     Finite optical depths cause a shift due to multiple scattering which is
     almost independent of the temperature.
} 
\end{figure}    

%  FIGURE 10
\begin{figure} 
\epsfig{file=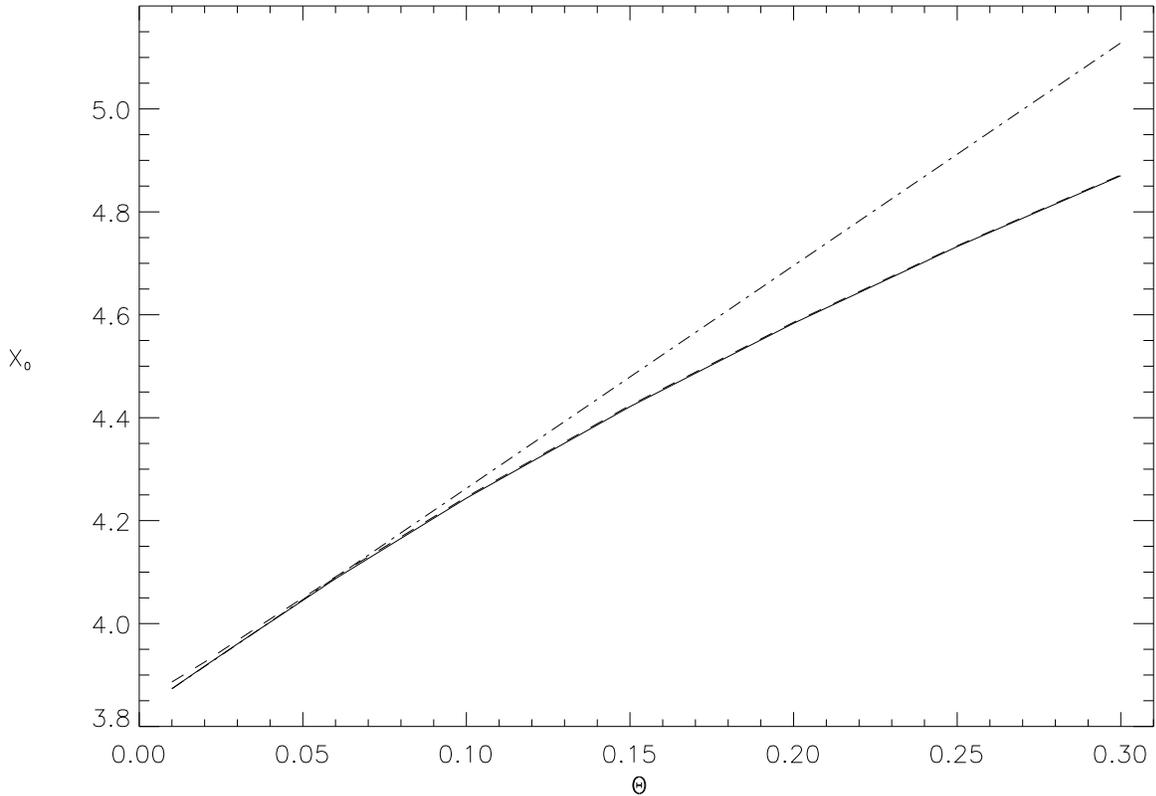,height=115mm}
\caption{\label{f:x0_MC_COLL}
     The cross-over frequency as a function of dimensionless temperature,
     $\Theta$, for our spherically symmetric model with infall. 
     The solid and dashed lines are for infall velocities $\beta_r = 0$ (reference static model)
     and $\beta_r = 0.05$.
     We used single scattering results to show only bulk motion effects. 
     The dashed dot line shows a linear approximation 
     (Challinor and Lasenby 1998a; Birkinshaw 1998).
     The effect of infall becomes inconsequential at high temperature since the 
     infall speed becomes negligible relative to the thermal speed of the electron.
} 
\end{figure} 

%  FIGURE 11
\begin{figure} 
\epsfig{file=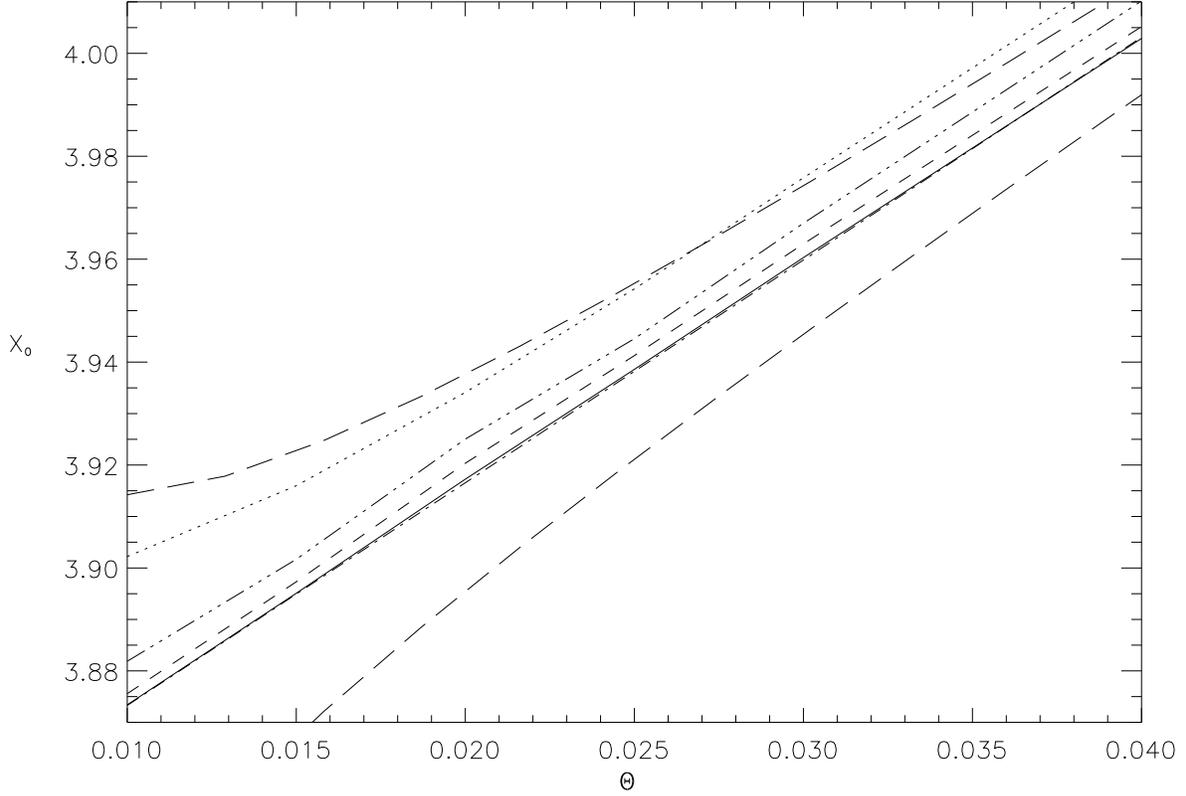,height=115mm}
\caption{ \label{f:x0_GLGX}
     The cross-over frequency as a function of dimensionless temperature,
     $\Theta$, for several optical depths in a static spherical plasma. 
     Our Monte Carlo results are from single scattering (solid line), 
     $\tau_0 = 0.02$ (short dashed line), 0.05 (dash, dot, dot, dot  line),
     and 0.1 (dotted line).
     The dash-dot line, which is hardly distinguishable from the solid line, 
     is the linear approximation of Challinor and Lasenby (1998a), and Birkinshaw (1998). 
     As a comparison, we plot the (large) effect of a radial cluster velocity of 
     $\pm$~100~km~s$^{-1}$ (kinematic effect) with long dashed lines 
     (Nozawa et al. 1998; Sazonov and Sunyaev 1998b).
} 
\end{figure} 

% % % % % % % % % % % % % % % % % % % % % % % % % % % % % % % % % % % % % % % % % % % % % % % % % % %

\end{document}